\documentclass[sigconf]{acmart}

\AtBeginDocument{%
  }

\acmConference[SBES '25]{Brazilian Symposium on Software Engineering}{September 22--26, 2025}{Recife, PE}
\setcopyright{none}
\renewcommand\footnotetextcopyrightpermission[1]{}

\usepackage{textcase}
\usepackage{pdfpages}   
\usepackage{caption}     
\usepackage{graphicx} 
\usepackage{listings}
\usepackage{xcolor}
\usepackage{booktabs}
\usepackage{adjustbox}
\usepackage{tabularx}
\usepackage{float} 
\usepackage{makecell}
\usepackage[most]{tcolorbox}
\usepackage{multirow}

\usepackage{microtype}
\usepackage{enumerate}

\lstdefinelanguage{Robot}{
    sensitive=false,
    morecomment=[l]{\#},
    morestring=[b]",
}

\lstdefinelanguage{json}{
  basicstyle=\ttfamily\tiny,
  numbers=none,
  numberstyle=\tiny\color{gray},
  stepnumber=1,
  numbersep=5pt,
  showstringspaces=false,
  breaklines=true,
  frame=single,
  backgroundcolor=\color{gray!7},
  literate=
   *{0}{{{\color{blue}0}}}{1}
    {1}{{{\color{blue}1}}}{1}
    {2}{{{\color{blue}2}}}{1}
    {3}{{{\color{blue}3}}}{1}
    {4}{{{\color{blue}4}}}{1}
    {5}{{{\color{blue}5}}}{1}
    {6}{{{\color{blue}6}}}{1}
    {7}{{{\color{blue}7}}}{1}
    {8}{{{\color{blue}8}}}{1}
    {9}{{{\color{blue}9}}}{1}
    {:}{{{\color{red}:}}}{1}
    {,}{{{\color{red},}}}{1}
    {"}{{{\color{red}"}}}{1}
    {{\{}}{{\{}}{1}
    {\}}{{\}}}{1}
}

\begin{document}

\newcommand{\genia}{GenIA-E2ETest}

\newcommand{\repogenia}{\url{https://github.com/uffsoftwaretesting/GenIA-E2ETest/}}

\newcommand{\repoAppCinema}{\url{https://github.com/elvisjuniorr/Projeto-Cinema}}

\title{GenIA-E2ETest: A Generative AI-Based Approach for End-to-End Test Automation}


\author{Elvis Júnior}
\affiliation{
  \institution{Universidade Federal Fluminense}
  \city{Niterói, RJ}
  \country{Brazil}}
\email{elvisjunior@id.uff.br}

\author{Alan Valejo}
\affiliation{%
  \institution{Universidade Federal de São Carlos}
  \city{São Carlos}
  \country{Brazil}}  
\email{alanvalejo@ufscar.br}

\author{Jorge Valverde-Rebaza}
\affiliation{%
  \institution{Tecnologico de Monterrey}
  \city{Mexico City}
  \country{Mexico}}  
\email{jvalverr@tec.mx}

\author{Vânia de Oliveira Neves}
\affiliation{%
  \institution{Universidade Federal Fluminense}
  \city{Niterói, RJ}
  \country{Brazil}}  
  \email{vania@ic.uff.br}

%\renewcommand{\shortauthors}{Neves and Garces}

\begin{abstract}
Software testing is essential to ensure system quality, but it remains time-consuming and error-prone when performed manually. Although recent advances in Large Language Models (LLMs) have enabled automated test generation, most existing solutions focus on unit testing and do not address the challenges of end-to-end (E2E) testing, which validates complete application workflows from user input to final system response. This paper introduces GenIA-E2ETest, which leverages generative AI to generate executable E2E test scripts from natural language descriptions automatically. We evaluated the approach on two web applications, assessing completeness, correctness, adaptation effort, and robustness. Results were encouraging: the scripts achieved an average of 77\% for
both element metrics, 82\% for precision of execution, 85\% for execution recall, required minimal manual adjustments (average manual modification rate of 10\%), and showed consistent performance in typical web scenarios. Although some sensitivity to context-dependent navigation and dynamic content was observed, the findings suggest that GenIA-E2ETest is a practical and effective solution to accelerate E2E test automation from natural language, reducing manual effort and broadening access to automated testing.
\end{abstract}


\maketitle

\section{\NoCaseChange{Introduction}}

% falar sobre engenharia de prompt, o que é, se está sendo muito utilizada.

% prompt engineering survey
% prompt engineering survey software engineering
% trabalhos futuros, usar outras estratégias de prompt ou outros LLM e outros datasets

% Contribuições:
% 1. Criação ou desemvolvimento de uma estratégia de pronpt multinível
% 2. Testar o desempenho de um LLM nesse cenário
% 3. Análise de desempenho em diferentes sistemas
% 4. disponibilização do prompt ou do projeto ou do sistema

Software testing plays a crucial role in ensuring the quality of complex applications. However, when performed manually, it remains a costly, time-consuming, and error-prone process. Despite the widespread availability of automation tools, many software products continue to suffer from insufficient testing and unsatisfactory quality~\cite{kochhar2013empirical}. Common limitations include the poor readability of generated tests~\cite{daka2015modeling, grano2019scented}, scalability and efficiency challenges, and reduced effectiveness, particularly in complex and dynamic scenarios.

Testing activities are traditionally organized into three levels: unit, integration, and system testing~\cite{delamaro2013introduccao}. While unit and integration tests are essential, system testing, and particularly End-to-End (E2E) testing, plays a critical role in validating complete user workflows under realistic conditions. E2E testing ensures not only the correctness of individual components but also their integrated operation from a user-centric perspective. However, implementing E2E tests is particularly challenging, requiring detailed mapping of user-interface elements, technical expertise with automation frameworks, and considerable time investment \cite{leotta2023challenges}. These factors hinder widespread adoption, especially among agile teams or organizations with limited technical or financial resources.

The advent of Large Language Models (LLMs), such as ChatGPT and Gemini, has opened new opportunities to automate test generation from natural language descriptions. However, most of the current applications of LLMs in software testing focus predominantly on low-level tasks, such as unit test generation~\cite{wang2024software, yang2024evaluation, wang2024hits, chen2024chatunitest}. While important, unit tests alone are insufficient to ensure the overall quality of complex systems. Fan et al.~\cite{wang2024software} emphasize the need to explore higher levels of testing, such as integration and system testing, as well as to develop strategies that can transform textual artifacts, such as user stories and test scenarios, into executable automated tests.

\begin{sloppypar}
Some commercial tools, such as Testim\footnote{\url{https://www.testim.io/}}, Functionize\footnote{\url{https://www.functionize.com/}}, and testRigor\footnote{\url{https://testrigor.com/}}, have started to leverage generative AI to simplify E2E testing. While promising, these solutions are proprietary, often impose high costs, and offer limited integration with widely adopted practices, such as supporting open frameworks like Robot Framework\footnote{\url{https://robotframework.org/}}. Moreover, they typically focus on regression testing and recorded workflows, restricting flexibility for specification-driven test creation.
\end{sloppypar}

\begin{sloppypar}
To address these challenges, we propose \genia, an open-source approach that transforms functional requirements, expressed in free-form natural language, into executable E2E test scripts for Robot Framework. \genia~is organized into three integrated modules: (i) \textbf{Scenario Modularization}, where a LLM parses textual descriptions to extract sequences of user actions and expected outcomes; (ii) \textbf{Extraction and Refinement of User Interface Elements}, where the application under test is crawled to automatically map user-interface components and their contextual attributes into a machine-readable catalog; and (iii) \textbf{Generation of Executable E2E Scripts}, where the interpreted requirements are combined with the mapped UI elements to generate E2E scripts compatible with existing automation frameworks (e.g., Robot Framework), with an emphasis on maintainability and readability.
\end{sloppypar}

The main contributions of this work are:

\begin{enumerate}[i] 
    \item the design of a multi-level prompting strategy to guide the LLM through structured requirement interpretation and test generation; 
    \item a comprehensive evaluation of the generated E2E test scripts, covering correctness, completeness, adaptation effort, and robustness across systems with varying complexity; 
    \item the release of a complete open-source tool that transforms functional requirements into executable E2E test scripts. The tool, along with all prompt templates and experimental artifacts, is publicly available, allowing researchers and practitioners to evaluate, reproduce, and extend our approach. Its modular design supports easy adaptation to different domains and testing environments, promoting practical adoption and encouraging future research in test automation driven by natural language requirements. \end{enumerate}

Our experimental evaluation demonstrated that \genia~is capable of generating executable and correct E2E test scripts with minimal manual intervention, particularly for web applications with conventional navigation flows. While challenges remain in scenarios involving dynamic content and context-dependent workflows, the results highlight the practical viability of integrating LLM-driven solutions into real-world automation pipelines.

%This paper is organized as follows...
\section{\NoCaseChange{E2E Test Automation in Web Applications}}

E2E testing is essential for verifying whether a web application functions correctly from the end-user’s perspective. These tests validate complete usage flows, covering multiple system layers, such as the user interface, business logic, and external integrations, by simulating real user interactions. Unlike unit tests, which focus on isolated components, or integration tests, which verify communication between modules, E2E tests ensure that the system as a whole behaves as expected~\cite{fowlerBroadStackTest}.

For instance, in an online shopping scenario, an E2E test might span from accessing the homepage to completing the payment, including steps such as product search, cart operations, and user authentication. Listing~\ref{lst:input_example} illustrates a manual test case example related to an invalid login scenario.

To support regression testing, a common recommendation is to automate tests that validate critical or frequently updated functionalities. This automation is often implemented using tools like Selenium\footnote{\url{https://www.selenium.dev/}}, which allows programmatic control of browsers and interaction with web interface elements. During this process, testers typically inspect each page’s HTML to locate relevant elements, such as input fields, buttons, and validation messages, and extract their selectors (e.g., XPaths or CSS selectors) to construct the test scripts.

Although Selenium can be used standalone, maintaining and understanding the resulting scripts can be challenging. For this reason, it is often combined with the Robot Framework\footnote{\url{https://robotframework.org/}}, an automation tool that structures tests using a descriptive, keyword-based syntax. This approach enhances readability, reusability, and accessibility, making it particularly useful for functional validation.

The integration between Robot and Selenium is provided by the \textit{SeleniumLibrary}\footnote{\url{https://robotframework.org/SeleniumLibrary/}}, a library that offers ready-to-use commands for typical user interactions, such as form filling, button clicking, or verifying messages on the page. This abstraction enables realistic simulations to be implemented in a modular, readable, and reproducible way. Listing~\ref{lst:robot_example} presents an automated test script created using Robot Framework and SeleniumLibrary, corresponding to the scenario introduced in Listing~\ref{lst:input_example}.

\section{\NoCaseChange{\genia: E2E Test Script Generation Through Multi-Level Prompting}}

While essential for quality assurance, E2E tests often pose challenges in terms of execution effort and automation complexity. These tests can be executed manually, by testers following step-by-step instructions, or automatically, using tools that simulate user interactions through the interface. However, manual execution tends to be repetitive, error-prone, and ineffective for regression testing, whereas automation typically demands significant technical expertise.

In industrial settings, it is common for teams to initially document test cases in natural language, executing them manually during early testing phases. As certain tests are recognized as critical or frequently executed, they become candidates for automation. Nevertheless, translating these scenarios into executable scripts remains a resource-intensive process, especially in agile environments with tight delivery schedules.

To address these challenges, we propose \genia, an approach for the automated generation of E2E test scripts from scenarios written in natural language. The approach combines semantic interpretation of functional requirements using LLMs with automated extraction of user interface elements, producing fully executable scripts compatible with the Robot Framework.

At the core of our solution lies a multi-level prompting strategy, in which distinct prompts are crafted to guide the LLM through three complementary tasks: (i) restructuring and modularizing test scenarios; (ii) identifying and refining user interface elements; and (iii) generating executable test scripts. Each level operates on a well-defined input and is designed to function independently, enabling a modular, extensible, and low-maintenance architecture. This strategy is conceptually inspired by AI Chains~\cite{wu2022ai}, which promote the decomposition of complex tasks into interpretable and controllable prompt steps. In the following sections, we describe each prompt level in detail.

%The proposed approach for automated End-to-End (E2E) test generation is designed as a three-phase pipeline that incrementally transforms high-level test scenarios into fully executable Robot Framework scripts. This pipeline integrates a Large Language Model (LLM), web scraping with Crawl4AI, and a modular architecture grounded in Behavior-Driven Development (BDD) principles. The primary goal is to provide a robust, scalable, and low-maintenance solution for test automation based on natural language test cases.

% \begin{figure*}[h]
%     \captionsetup{position=above}
%     \centering
%     \caption{Overview of the \genia~Approach and Multi-Level Prompting Strategy}
%     \includegraphics[width=\textwidth, height=0.6\textheight, keepaspectratio]{genia.png}
%     \label{fig:genia_approach}
% \end{figure*}

\begin{figure}[h]
    \captionsetup{position=above}
    \centering
    \caption{Overview of the \genia~approach and multi-level prompting strategy}
    \includegraphics[width=0.5\textwidth, keepaspectratio]{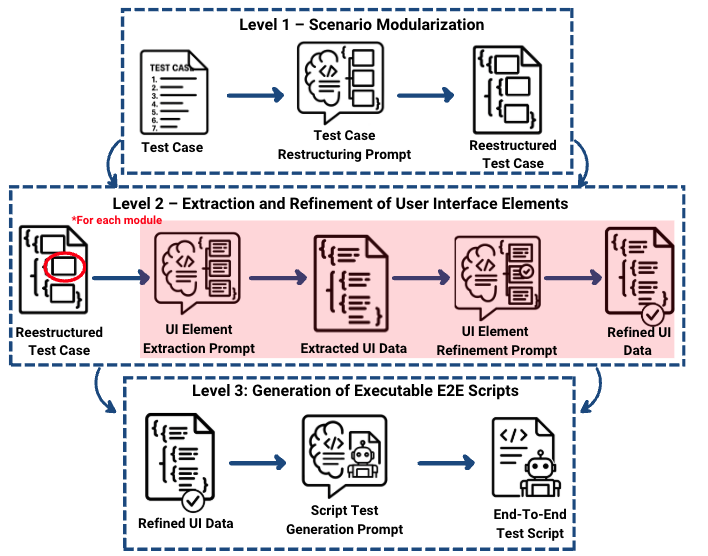}
    \label{fig:genia_approach}
\end{figure}

\vspace{-0.8em}
\subsection{Level 1 – Scenario Modularization} \label{sec:level1}

%The first phase focuses on restructuring raw test cases—typically written in Gherkin syntax or plain natural language—into a structured and modular JSON format. This transformation is initiated by a prompt to an LLM via the OpenAI API.

The first prompt level is responsible for transforming unstructured or semi-structured test scenarios ,typically written in natural language or step-by-step procedural format, into a structured and modular JSON representation. This transformation is carried out by submitting a structured prompt to a generative LLM via an API.

%The LLM is instructed to parse the input and divide it into logical modules, each representing a distinct web page within the user journey. Each module is anchored by its respective URL and includes a chronologically ordered list of \texttt{execution\_steps}, such as form inputs, button clicks, or assertions.

The prompt was designed based on prompt engineering best practices, as recommended by White et al.~\cite{white2023prompt}, including the explicit definition of objectives, the specification of the expected output structure, and the use of personas to guide the model's behavior. In this case, the LLM is instructed to act as a highly skilled software test automation engineer responsible for segmenting the test case into logical modules. Each module corresponds to a single application page, identified by its full URL. It contains a chronologically ordered list of user actions \texttt{execution\_steps}, such as filling out fields, clicking buttons, or performing assertions. 
This approach enables the LLM to produce a predictable and structured output even without explicit examples in the prompt, characterizing a \textit{zero-shot prompting strategy}. This prompting strategy is particularly suitable for evaluating the model’s ability to generalize to tasks not explicitly seen during training, and it is commonly adopted in software testing applications~\cite{wang2024software}.

\begin{lstlisting}[caption={Manual test scenario: natural language steps with navigation URLs and validation actions},
label={lst:input_example}]
urls = ["http://automationexercise.com","https://automationexercise.com/login"]
Test Case 1: Login User with incorrect email and password
1. Launch browser and navigate to url 'http://automationexercise.com'
2. Click on 'Signup / Login' button
3. Enter incorrect email address and password
4. Click 'login' button
5. Verify error 'Your email or password is incorrect!' is visible
\end{lstlisting}
%Any step that leads to a change in the web page (i.e., URL transition) marks the start of a new module, supporting a page-aware design. This modular structure improves traceability and paves the way for more efficient automation.

Any action that results in a URL transition is treated as a module boundary, marking the end of the current module and the beginning of the next. For instance, consider an invalid login scenario in which the user accesses the login page, fills in incorrect credentials, and clicks the ``Login'' button. Upon submission, the system redirects the user to a dedicated error page that informs them of the failed authentication attempt. In this case, all actions performed up to the click on ``Login'' are grouped within the login page module, while the error message validation on the redirected page belongs to the error page module. This modular structure improves traceability and paves the way for more efficient automation.

%At this point, the \texttt{extracted\_data} field in each step is left empty, as no HTML analysis has been performed yet.

\begin{lstlisting}[language=json, caption={Test case in JSON format: execution steps, navigation URLs, page modules, and validation details},
label={lst:output_example},
  numbers=left,    
  numberstyle=\tiny\color{gray}, 
  numbersep=10pt ]
  "testCase": "Login User with incorrect email and password",
  "modules": [
      {
      "url": "http://automationexercise.com",
      "purpose": "Home page of the application",
      "execution_steps": [{
          "step": "Launch browser and navigate to url 'http://automationexercise.com'",
          "extracted_data": []
        },{
          "step": "Click on 'Signup / Login' button",
          "extracted_data": []
        }]},
      {
      "url": "https://automationexercise.com/login",
      "purpose": "Login page for users to enter their credentials",
      "execution_steps": [{
          "step": "Enter incorrect email address and password",
          "extracted_data": []
        },{
          "step": "Click 'login' button",
          "extracted_data": []
        },{
          "step": "Verify error 'Your email or password is incorrect!' is visible",
          "extracted_data": []
        } ] } ]
\end{lstlisting}

In this process, the prompt receives as input a test case described in natural language, as illustrated in Listing~\ref{lst:input_example}, where the expected user actions and visited URLs are specified. Based on this input, the LLM generates a structured JSON output, shown in Listing~\ref{lst:output_example}, which includes the identified modules, their associated URLs, and the extracted execution steps. The \texttt{extracted\_data} field remains empty at this stage, as no HTML interface analysis has been performed yet. Level 1 in Figure~\ref{fig:genia_approach} illustrates this stage of the approach.

\vspace{-0.8em}
\subsection{Level 2 – Extraction and Refinement of User Interface Elements} \label{sec:level2}

%This second phase enriches the test structure by automatically identifying and refining the web elements required for automation within each module.

Once the test scenarios have been divided into modules, each representing a page in the user's navigation flow, in Level 1 (Section \ref{sec:level1}), Level 2 of our multi-level prompting strategy aims to record, for each test step, the interface elements required for its automation. To this end, relevant HTML elements are identified and assigned to the \texttt{extracted\_data} field, allowing each user action to be associated with its corresponding UI selectors, which will later be used to generate and execute automated test scripts.

This level involves two distinct prompts: one responsible for the initial extraction of UI elements (Section \ref{sec:level2-1}) and a second, applied afterward, to refine and validate the extracted elements (Section \ref{sec:level2-2}). The second prompt is necessary because the LLM may occasionally generate selectors that are incorrect, incomplete, or redundant. This additional step thus serves as a verification and enhancement mechanism, contributing to the robustness and reliability of the extracted data. Although they operate independently, the two prompts form a complementary strategy with a single goal: to accurately and semantically map UI elements, reducing inconsistencies and improving the overall quality of the generated test scripts. The sequential use of prompts at this stage is conceptually aligned with the principles of the AI Chains model~\cite{wu2022ai}, as previously discussed.

This prompt orchestration also brings practical architectural benefits. By restricting the LLM’s scope to a single page per module, the approach improves analysis precision, prevents excessive data accumulation, and reduces memory usage. Furthermore, its modular design supports the independent evolution of extraction and refinement logic, allowing future improvements without changes to the overall iteration structure.

\subsubsection{UI Element Extraction Prompt} \label{sec:level2-1}

%A single \texttt{for} loop is responsible for iterating through each module, processing one web page at a time in isolation—just as a user would interact with the application step by step.

Each module produced in the output of Level 1 is processed in isolation, simulating a page-by-page navigation flow, similar to how a tester would interact with the application manually. For each module, the corresponding page's HTML content is retrieved using \textit{Crawl4AI}\footnote{https://github.com/unclecode/crawl4ai}, an open-source tool capable of capturing complete and dynamic DOM representations.

%\textbf{Step 1: Extraction using Crawl4AI and LLM:} \\
%For each module:
%\begin{itemize}
%    \item The corresponding web page’s HTML is retrieved using Crawl4AI, an open-source web scraping tool capable of extracting dynamic content.
%    \item The retrieved HTML and the module’s test steps are sent to the LLM.
%    \item The LLM returns a structured list of elements relevant to the test case, with each element containing:
 %   \begin{itemize}
 %       \item \texttt{type} – The HTML element type (e.g., \textit{input}, \textit{button}, \textit{select}).
%        \item \texttt{request\_description} – A natural language explanation of what the element expects from the user (e.g., "Enter with the First Name").
%        \item \texttt{identifier\_type} – The method used to locate the element (preferably \textit{XPath}, but can be \textit{id}, \textit{name}, etc.).
%        \item \texttt{identifier\_tracking} – The exact identifier used to locate the element, such as a full XPath (e.g., \texttt{//*[@id="form"]
%        /input[1]}).
%    \end{itemize}
%    \item These results are stored in the \texttt{extracted\_data} field of the module.
%\end{itemize}

This HTML, along with the list of \texttt{execution\_steps}, is then submitted to an LLM via a structured prompt. In this prompt, the LLM is instructed to act as a test automation manager and extract only the elements required to perform each step. The expected output is a list of relevant UI elements, each annotated with: (i) the type (e.g., \textit{input}, \textit{button}); (ii) a natural language description of the expected user interaction (e.g., ``Enter with the First Name''); (iii) the locator strategy (preferably XPath); (iv) the full selector expression used to identify the element; and (v) the name of the corresponding test step. This prompt follows a \textit{zero-shot prompting strategy}, relying solely on clear instructions and a well-defined output schema.

The extracted elements are stored in the \texttt{extracted\_data} field of the corresponding module. An example of the resulting JSON structure is shown in Listing~\ref{lst:extract_example}, where the step ``Click on 'Signup / Login' button'', defined in lines 9–12 of Listing~\ref{lst:output_example}, was enriched with the data extracted by the LLM.

\begin{lstlisting}[language=json, caption={Populated JSON output with UI mappings and execution context},
label={lst:extract_example},
  numbers=left, 
  firstnumber=11,
  numberstyle=\tiny\color{gray}, 
  numbersep=10pt ]]
          "step": "Click on 'Signup / Login' button",
          "extracted_data": [{
              "type": "button",
              "request_description": "Button to navigate to the Signup / Login page",
              "identifier_type": "XPath",
              "identifier_tracking": "//a[contains(text(), 'Signup / Login')]"
            },{
              "type": "button",
              "request_description": "Button to navigate to the Signup / Login page",
              "identifier_type": "XPath",
              "identifier_tracking": "//*[@id='header']/div[2]/div/div/div[2]/div[1]/ul/li[1]/a"
            }] ...  
\end{lstlisting}

\subsubsection{UI Element Refinement Prompt} \label{sec:level2-2}

%Still within the same loop iteration, a second LLM prompt refines the extracted elements:
%\begin{itemize}
%    \item Removes duplicates or irrelevant elements
%    \item Improves descriptions for clarity
%    \item Enhances XPath robustness
%\end{itemize}
Immediately after extracting each module, a second prompt is used to refine the collected elements. At this stage, the LLM assumes the role of a senior E2E test engineer, responsible for reviewing the quality of the extracted data and optimizing the identified selectors. As discussed earlier, this additional step serves as a validation mechanism, ensuring that the extraction performed by the prompt described in Section~\ref{sec:level2-1} has produced elements that are correct, complete, and suitable for automation.

The LLM is instructed to remove duplicate or irrelevant elements, validate and improve the selectors (with emphasis on XPath-based ones), and review the descriptions and their associations with test steps for clarity and consistency. As with the extraction prompt, this refinement step also follows a \textit{zero-shot prompting} strategy, relying on clear instructions and a well-defined output structure without the inclusion of explicit examples.

To better illustrate the impact of the refinement prompt, Listing~\ref{lst:revised_example} presents an excerpt of the refined JSON structure. The original listing maps two distinct XPath expressions to the step ``Click on 'Signup / Login' button'' (lines 14 and 19), resulting in redundant selectors for the same UI element. The revised version, by contrast, consolidates this into a single, unambiguous mapping, thereby eliminating redundancy and improving both clarity and maintainability of the test specification.

%This immediate refinement avoids an additional processing pass and maintains architectural simplicity.

%\textbf{Benefits of Module-Based Sequential Processing}

%This strategy offers several advantages:
%\begin{itemize}
%    \item \textbf{Scoped Context:} Keeps LLM focus on one page at a time, enhancing precision
%    \item \textbf{Lower Memory Overhead:} Reduces data accumulation and memory usage
%    \item \textbf{Modularity:} Enables future enhancements to extraction/refinement logic without altering the loop structure
%\end{itemize}

\begin{lstlisting}[language=json, caption={Refined JSON with precise UI element mappings, optimized for E2E test execution.},
label={lst:revised_example},
  numbers=left,    
  numberstyle=\tiny\color{gray}, 
  firstnumber=11,
  numbersep=10pt ]]
          "step": "Click on 'Signup / Login' button",
          "extracted_data": [{
              "type": "button",
              "request_description": "Button to navigate to the Signup / Login page",
              "identifier_type": "XPath",
              "identifier_tracking": "//a[contains(text(), 'Signup / Login')]"
            }] ...
\end{lstlisting}

This immediate refinement eliminates the need for a subsequent verification stage and contributes to the architectural simplicity of the approach. Ultimately, it ensures that the data is properly prepared for test script generation, with robust, accurate, and semantically meaningful UI element mappings. The combination of extraction and refinement defines Level 2 of \genia, which is also represented as Level 2 in Figure~\ref{fig:genia_approach}.

\subsection{Level 3: Generation of Executable E2E Scripts}\label{sec:level3}

%In the final phase, the \texttt{RefinedExtractedData.json} is used to generate a complete Robot Framework test script. A prompt is sent to the LLM instructing it to simulate the reasoning of a skilled QA engineer.

In Level 3 of \genia, the validated test specification produced in the previous levels, structured as a JSON file, is used to generate a complete end-to-end automated test script in an executable format. This level corresponds to Level 3 in Figure~\ref{fig:genia_approach}, which represents the final stage of the multi-level prompting strategy. Although the approach is agnostic to the target test automation framework, the current implementation uses the Robot Framework for script generation and execution.

A final prompt (fourth in total) is submitted to the LLM, instructing it to act as an experienced test automation engineer. Based on the modular JSON structure generated in the earlier levels, the model is expected to produce a technically correct, executable, readable, and maintainable script. In the current implementation, the prompt is specifically designed to guide the model in generating scripts strictly following the Python-based Robot Framework syntax and using the Selenium library. As in previous levels, this prompt follows a \textit{zero-shot} strategy, relying on clear instructions and a well-defined output structure without explicit examples.

%The resulting script:
%\begin{itemize}
%    \item Uses \texttt{SeleniumLibrary} and Python
%    \item Includes sections: \texttt{*** Settings ***}, \texttt{*** Variables ***}, \texttt{*** Test Cases ***}, \texttt{*** Keywords ***}
%    \item Is executable without manual editing
%\end{itemize}

%The final output, encoded in UTF-8, is a ready-to-run, human-readable automation script—completing the bridge from natural language to technical implementation.

Listing~\ref{lst:robot_example} shows an example of the generated test script based on the JSON file from Listing~\ref{lst:revised_example}, corresponding to the test scenario described in Listing~\ref{lst:input_example}. The final output is a ready-to-run automation script, completing the transformation from natural language scenarios into operational technical scripts -- thus bridging the gap between requirements specification and automated test execution.

Although the current prompt is tailored for the Python-based Robot Framework and Selenium, the overall structure of the test specification, as well as the multi-level prompting strategy, is agnostic to specific automation tools. With minimal adjustments to the prompt instructions and output formatting, the same approach can be adapted to generate test scripts for other technologies, such as Cypress, Playwright, or JUnit, thus demonstrating the extensibility of \genia~to different testing stacks. All prompts used in this study are available in our public repository\footnote{\repogenia}, supporting reuse and adaptation to different test automation stacks.

\begin{lstlisting}[caption=Fully structured Robot Framework script generated from validated JSON data,
label={lst:robot_example}]
*** Settings ***
Library           SeleniumLibrary

*** Variables ***
${URL}            http://automationexercise.com
${LOGIN_URL}     https://automationexercise.com/login
${INCORRECT_EMAIL}    test@example.com
${INCORRECT_PASSWORD}     wrongpassword

*** Test Cases ***
Login User with Incorrect Email and Password
    Open Browser    ${URL}    chrome
    Maximize Browser Window
    Click Element    //a[contains(text(), 'Signup / Login')]
    Input Text    //*[@id='form']//input[@name='email']    ${INCORRECT_EMAIL}
    Input Text    //*[@id='form']//input[@name='password']    ${INCORRECT_PASSWORD}
    Click Button    //*[@id='form']//button[@type='submit']
    Element Should Be Visible    //div[contains(text(), 'Your email or password is incorrect!')]  
    Close Browser
\end{lstlisting}

\section{\NoCaseChange{Experimental Evaluation}}

This section presents the empirical evaluation of the \genia~ approach. The study follows the principles of Engineering Research, in accordance with the standards established by Ralph et al.~\cite{ralph2020empirical}. This type of investigation is appropriate when the goal is to propose, implement, and evaluate innovative technical solutions to practical problems in Software Engineering.

To assess its performance, the solution was applied to two web applications, aiming to investigate its correctness, completeness, adaptation effort, and robustness across applications with different complexity levels. To guide the evaluation, we adopted the Goal-Question-Metric (GQM) model~\cite{caldiera1994goal}, ensuring alignment between the study’s objectives, research questions, and the metrics used for data collection and analysis. 

\subsection{Research Design} \label{sec:res_design}

Following the GQM model, the goal of our study is formally defined as follows:

\begin{tcolorbox}[colback=gray!5!white,colframe=black!75!black]
\textbf{Analyze} the \genia~approach \textbf{with the purpose of} evaluating its correctness, completeness, adaptation effort, and robustness, \textbf{from the perspective of} software testers and researchers,\textbf{ in the context of} web applications.
\end{tcolorbox}

%This section outlines the experimental setup used to evaluate the effectiveness of the proposed AI-driven approach for automated end-to-end test case generation. The goal is to assess its correctness, completeness, and feasibility, based on a structured analysis of generated test scripts applied to two distinct applications with varying degrees of complexity.

Based on this goal, we formulated four research questions and defined a set of quantitative metrics to evaluate them, as described below.

    \vspace{-0.85em}
\subsubsection{Research Questions and Metrics}

\noindent\textbf{RQ1 – \underline{Correctness:}
  To what extent are the generated test scripts correct with respect to the identification of interface elements and the execution of the expected actions? }  \\
    \textit{Rationale:} Correctness is an essential requirement for test scripts to be used in real automation pipelines. This question investigates two complementary aspects: (i) whether the interface elements were correctly identified from the test specification, as defined in Level 2 of \genia~(Section~\ref{sec:level2}), and (ii) whether the generated scripts are technically valid, executable, and behave as expected, as defined in Level 3 (Section~\ref{sec:level3}).\\
    \textit{Metrics used:} Precision of Element Generation (C/G), Element Recall (C/E), Precision of Execution (CS/GS), Execution Recall (CS/ES) — see Table~\ref{tab:metricas}. \\
    
    \noindent \textbf{RQ2 – \underline{Completeness:}  
    Do the generated scripts cover all the elements and steps defined in the test scenarios? }  \\
    \textit{Rationale:} In addition to being correct, test scripts must be complete to ensure full verification of the intended scenario. This question assesses the coverage of both interface elements and expected actions in relation to the target scenario.  \\ 
    \textit{Metrics used:} Element Coverage (G/E) and Step Coverage (GS/ES) — see Table~\ref{tab:metricas}. \\
    
   \noindent\  \textbf{RQ3 – \underline{Adaptation Effort:}  
What is the manual effort required to adapt the generated test scripts to make them executable?} \\
\textit{Rationale:}  
Despite advances in large language models, automatically generated test scripts often require minor adjustments to run correctly. This question focuses on the initial manual effort required to make the generated scripts executable, which is a relevant concern in real-world testing pipelines. The adaptations may include fixing syntax issues, updating locators, or correcting step sequences. This aspect is conceptually distinct from long-term maintainability and is evaluated immediately after script generation.  \\
\textit{Metrics used:} Modified Lines (M), Lines of Code (LOC), and Manual Modification Rate (MR) — see Table~\ref{tab:metricas}. \\
    
\noindent\     \textbf{RQ4 – \underline{Robustness Across Contexts:}  
    How consistent is the performance of the approach across different types of interaction and system contexts?} \\    
    \textit{Rationale:}  
        A robust test generation approach should produce reliable results when applied to systems with different structures, interface designs, and interaction models. This question evaluates the consistency of script generation and execution outcomes when \genia~ is applied to two distinct web applications.  \\  
    \textit{Metrics used:} All metrics used in RQ1 and RQ2.

\begin{table*}[ht]
\centering
% \tiny
\scriptsize
\caption{Metrics used in the experimental evaluation}
\label{tab:metricas}
\begin{tabular}{p{4cm} p{1.5cm} p{10cm}}
\toprule
\textbf{Metric Name} & \textbf{Formula} & \textbf{Description} \\
\midrule
Expected Elements (E) & -- & Total number of UI elements that should appear in the script, based on the test scenario. \\
Generated Elements (G) & -- & Total number of elements included by the model in the generated script. \\
Correct Elements (C) & -- & Number of generated elements that match the expected type, role, and locator. \\
\addlinespace
Element Coverage (\%) & (G / E) × 100 & Proportion of expected elements included in the generated script. \\
Precision of Element Generation (\%) & (C / G) × 100 & Proportion of generated elements that were correctly identified. \\
Element Recall (\%) & (C / E) × 100 & Proportion of expected elements that were correctly retrieved. \\
\addlinespace
Expected Steps (ES) & -- & Total number of steps expected in the test scenario (e.g., click, fill, assert). \\
Generated Steps (GS) & -- & Total number of steps included in the generated script. \\
Correct Steps (CS) & -- & Steps that were successfully executed by the script. \\
\addlinespace
Step Coverage (\%) & (GS / ES) × 100& Proportion of expected steps included in the generated script. \\
Precision of Execution (\%) & (CS / GS) × 100 & Proportion of generated steps that executed successfully. \\
Execution Recall (\%) & (CS / ES) × 100 & Proportion of all expected steps that executed correctly. \\
\addlinespace
Lines of Code (LOC) & -- & Total number of lines in the generated script. \\
Modified Lines (ML) & -- & Number of lines manually edited to enable execution. \\
Manual Modification Rate (MR) (\%) & (ML / LOC) × 100 & Proportion of lines that required manual adjustment. \\
\bottomrule
\end{tabular}
\end{table*}

\subsubsection{Systems Under Evaluation} 
%To evaluate the generalizability and robustness of the solution, two applications were chosen:

To answer the research questions, we selected two web applications with distinct characteristics, aiming to represent different interaction and structural profiles: \\

\textbf{WebApp 1 – AutomationExercise}\footnote{\url{https://automationexercise.com}} \\ AutomationExercise is a publicly available website widely used for instructional purposes and test automation practice. It provides typical web functionalities such as user login, registration, form validation, product search, and shopping cart interactions. In addition, the platform offers a curated list of test case scenarios intended for practice and evaluation purposes\footnote{\url{https://www.automationexercise.com/test_cases}}, which facilitates script generation, benchmarking, and reproducibility. While not designed as a research-grade benchmark, its consistent structure and wide range of interactive elements make it a suitable environment for evaluating the generation of functional end-to-end test scripts.

\textbf{WebApp 2 – Movie Ticketing Web App} \footnote{Repository: \repoAppCinema} \\
This system was developed by the first author using React and Vite and is locally hosted to ensure full control over its source code and execution environment. It simulates a complete movie ticket booking workflow, including user login and registration, movie catalog navigation, session selection, form submission, and administrative features. Its structured navigation, dynamic routing, and use of modern frontend components make it suitable for evaluating AI-based E2E test generation in more complex and stateful interaction scenarios. Additionally, because this system was developed independently and was not publicly available during the training of current LLMs, it offers a controlled evaluation environment that helps mitigate potential bias or memorization effects often associated with publicly known applications.

\subsubsection{Test Case Selection and Diversity}

To evaluate the approach in different interaction contexts, twelve test cases were selected, six per application, based on their functional diversity and representativeness. Rather than aiming for exhaustive coverage of each system, we prioritized scenarios that reflect realistic usage flows and recurring challenges in end-to-end (E2E) test automation. The choice of a reduced sample size was motivated by the need to conduct an in-depth analysis of the generated scripts' quality, enabling the assessment of the approach’s applicability across different levels of functional complexity with control and analytical rigor. All generated scripts were manually reviewed and executed to ensure accurate evaluation of their correctness, completeness, and required adaptation effort.

To reflect a wide range of real-world user interactions, the twelve selected test cases cover different goals and usage patterns, including successful and unsuccessful login attempts, form submissions, user registrations, subscription verifications, and search and filtering actions. The scenarios also explore dynamic interface behaviors such as field validations, error messages after invalid inputs, page scrolling to locate elements in different sections, and file uploads in contact forms. Listing~\ref{lst:input_example} illustrates one of the evaluated cases, corresponding to WebApp1-TC3, which assessed the system's behavior when rejecting invalid login credentials. WebApp1-TC6, in turn, involved submitting a form with a file upload and verifying redirection to the home page. All complete scenarios are available in our repository\footnote{\repogenia}, enabling reproducibility.

\begin{table}[ht]
\centering
\caption{Test cases used in the evaluation}
\label{tab:test-case-info}
\scriptsize
\begin{tabularx}{\columnwidth}{l l X}
\toprule
\textbf{Application} & \textbf{Test Case ID} & \textbf{Title} \\
\midrule
\multirow{6}{*}{WebApp 1} 
  & TC1 & Verify Subscription in Home Page \\
  & TC2 & Verify Scroll Up Using 'Arrow' Button \\
  & TC3 & Login User with Incorrect Email and Password \\
  & TC4 & Verify Subscription in Cart Page \\
  & TC5 & Register User \\
  & TC6 & Contact Us Form \\
\midrule
\multirow{6}{*}{WebApp 2} 
  & TC1 & Successfully Search for a Movie \\
  & TC2 & Successfully Filter Movies by State \\
  & TC3 & Unsuccessfully Login with Incorrect Credentials \\
  & TC4 & Successfully Navigate to Movie Details Page \\
  & TC5 & Successfully Register a New User \\
  & TC6 & Successfully Register a New Movie \\
\bottomrule
\end{tabularx}
\end{table}

\begin{comment}

\begin{table}[ht]
\centering
\caption{Test Case Information}
\label{tab:test-case-info}
\scriptsize
\begin{tabularx}{\columnwidth}{c >{\centering\arraybackslash}p{0.5\columnwidth} c}
\toprule
\textbf{Test Case} & \textbf{Title} & \textbf{Level} \\
\midrule
WebApp1-TC1 & Verify Subscription in home page & EASY \\
WebApp1-TC2 & Verify Scroll Up using 'Arrow' button & EASY \\
WebApp1-TC3 & Login User with incorrect email and password & INTERMEDIATE \\
WebApp1-TC4 & Verify Subscription in Cart page & INTERMEDIATE \\
WebApp1-TC5 & Register User & COMPLEX \\
WebApp1-TC6 & Contact Us Form & COMPLEX \\
WebApp2-TC1 & Successfully Search for a Movie & EASY \\
WebApp2-TC2 & Successfully Filter Movies by State & EASY \\
WebApp2-TC3 & Unsuccessfully Login with Incorrect Credentials & INTERMEDIATE \\
WebApp2-TC4 & Successfully Navigate to Movie Details Page & INTERMEDIATE \\
WebApp2-TC5 & Successfully Register a New User & COMPLEX \\
WebApp2-TC6 & Successfully Register a New Movie & COMPLEX \\
\bottomrule
\end{tabularx}
\end{table}

\end{comment}

\vspace{-0.85em}
\subsubsection{Execution}

The experimental procedure followed a well- defined structure to ensure consistency and reproducibility. All test scripts were generated using the ChatGPT-4o model (\textit{gpt-4o-mini}), accessed via the OpenAI API. To enforce determinism in the model's outputs, the temperature parameter was set to 0 for all executions, thereby eliminating stochastic variation in script generation. Nevertheless, each test case listed in Table~\ref{tab:test-case-info} was executed three times using the same prompts and configurations. This repetition aimed to mitigate potential variations during the test execution phase. In total, 36 executions were performed, contributing to a more reliable evaluation of \genia~approach.

All executions were performed on a machine running Windows 11 Home Single Language 64-bit (Build 22631), equipped with an Intel(R) Core(TM) i7-1165G7 CPU at 2.80GHz (8 cores) and 12GB of RAM. The environment included Google Chrome version 135 for test execution, Node.js v23.11.0 with npm v10.9.2 for application setup, and Python v3.12.3 with Robot Framework v7.2.2 for test automation. The Crawl4AI tool (version 0.5.0.post8) was used to assist in user interface element extraction. The total time required to generate all test scripts was approximately 26 minutes.

To ensure execution isolation, browser sessions were reset between test runs by clearing cache, localStorage, and cookies. All tests were conducted in a clean environment, free from residual data or side effects from previous executions.

The scripts were executed using Robot Framework integrated with Selenium WebDriver. Each generated script was manually and individually executed under supervision to ensure direct observation of its runtime behavior.

\subsubsection{Data Analysis Procedure}

After execution, each generated test script was manually reviewed and executed to verify its behavior against the expected scenario. The evaluation process was conducted individually for each script, and the outcomes were systematically recorded in a structured results table.

For scripts that did not execute correctly, minimal manual adjustments were performed to make them executable. During this process, the effort required for each correction was documented, including the number of modified lines and the estimated time spent on the adjustment.

Once all executions and fixes were completed, the collected data were analyzed based on the metrics defined in Table~\ref{tab:metricas}, enabling the assessment of correctness, completeness, and adaptation effort. The results of this analysis are presented in the next section.

\vspace{-0.8em}
\subsection{Results}
This section presents the results of the empirical evaluation conducted with the \genia~approach, guided by the GQM model (Section~\ref{sec:res_design}). We assessed four quality attributes: \textit{correctness}, \textit{completeness}, \textit{adaptation effort}, and \textit{robustness}.

The evaluation included 36 script executions (12 test cases, each executed three times), manually supervised using Robot Framework and Selenium WebDriver to observe runtime behavior. The complete adaptation and supervised execution of all 36 scripts required approximately 3 hours, in addition to 26 minutes for prompt-based script generation.

Table~\ref{tab:analysis-data} summarizes the aggregated metrics. Results are presented by research question, highlighting patterns, outliers, and implications.

\begin{table*}[ht]
\centering
\setlength{\tabcolsep}{1pt} % Ajuste esse valor conforme necessário
\caption{Summary of aggregated and averaged results for the defined metrics}
\label{tab:analysis-data}
\scriptsize
\begin{tabularx}{\textwidth}{>{\centering\arraybackslash}X 
                              >{\centering\arraybackslash}p{1.5cm} 
                              >{\centering\arraybackslash}X 
                              >{\centering\arraybackslash}X 
                              >{\centering\arraybackslash}X 
                              >{\centering\arraybackslash}X 
                              >{\centering\arraybackslash}X 
                              >{\centering\arraybackslash}X 
                              >{\centering\arraybackslash}X 
                              >{\centering\arraybackslash}X 
                              >{\centering\arraybackslash}X 
                              >{\centering\arraybackslash}X 
                              >{\centering\arraybackslash}X 
                              >{\centering\arraybackslash}X 
                              >{\centering\arraybackslash}X
                              >{\centering\arraybackslash}X
                              }
\toprule
\footnotesize\textbf{TC} & 
\footnotesize\textbf{E} & 
\footnotesize\textbf{G} & 
\footnotesize\textbf{C} & 
\footnotesize\textbf{G/E} & 
\footnotesize\textbf{C/G} & 
\footnotesize\textbf{C/E} & 
\footnotesize\textbf{LOC} & 
\footnotesize\textbf{M} & 
\footnotesize\textbf{MR} & 
\footnotesize\textbf{ES} & 
\footnotesize\textbf{GE} & 
\footnotesize\textbf{CS} & 
\footnotesize\textbf{GS/ES} & 
\footnotesize\textbf{CS/GS} & 
\footnotesize\textbf{CS/ES} \\

\midrule
\texttt{WebApp1-TC1} & 14 & 14 & 14 & 100\% & 100\% & 100\% & 64 & 3 & 5\% & 32 & 32 & 29 & 100\% & 91\% & 91\% \\
\texttt{WebApp1-TC2} & 23 & 23 & 21 & 100\% & 91\% & 91\% & 77 & 5 & 6\% & 32 & 41 & 27 & 128\% & 66\% & 84\% \\
\texttt{WebApp1-TC3} & 17 & 17 & 14 & 100\% & 82\% & 82\% & 67 & 4 & 6\% & 34 & 34 & 31 & 100\% & 91\% & 91\% \\
\texttt{WebApp1-TC4} & 15 & 15 & 15 & 100\% & 100\% & 100\% & 64 & 4 & 6\% & 29 & 31 & 27 & 107\% & 87\% & 93\% \\
\texttt{WebApp1-TC5} & 94 & 94 & 11 & 100\% & 12\% & 12\% & 171 & 83 & 49\% & 108 & 108 & 26 & 100\% & 24\% & 24\% \\
\texttt{WebApp1-TC6} & 39 & 39 & 34 & 100\% & 87\% & 87\% & 95 & 7 & 7\% & 56 & 58 & 51 & 104\% & 88\% & 91\% \\
\texttt{WebApp2-TC1} & 6 & 6 & 3 & 100\% & 50\% & 50\% & 49 & 3 & 6\% & 21 & 21 & 21 & 100\% & 100\% & 100\% \\
\texttt{WebApp2-TC2} & 9 & 9 & 9 & 100\% & 100\% & 100\% & 48 & 1 & 2\% & 21 & 21 & 20 & 100\% & 95\% & 95\% \\
\texttt{WebApp2-TC3} & 17 & 17 & 14 & 100\% & 82\% & 82\% & 56 & 3 & 5\% & 26 & 26 & 23 & 100\% & 88\% & 88\% \\
\texttt{WebApp2-TC4} & 10 & 10 & 10 & 100\% & 100\% & 100\% & 42 & 0 & 0\% & 19 & 19 & 19 & 100\% & 100\% & 100\% \\
\texttt{WebApp2-TC5} & 35 & 35 & 18 & 100\% & 51\% & 51\% & 95 & 9 & 9\% & 47 & 49 & 42 & 104\% & 86\% & 89\% \\
\texttt{WebApp2-TC6} & 34 & 34 & 24 & 100\% & 71\% & 71\% & 77 & 12 & 16\% & 45 & 45 & 33 & 100\% & 73\% & 73\% \\
\midrule
\textbf{General} & 313 & 313 & 187 & 100\% & 77\% & 77\% & 905 & 134 & 10\% & 470 & 485 & 349 & 104\% & 82\% & 85\% \\
\bottomrule
\end{tabularx}
\end{table*}

\subsubsection{RQ1 – Correctness} \label{sec:rq1}

To assess correctness, we analyzed the accuracy of interface element identification and the successful execution of test steps using four metrics: \textit{Precision of Element Generation} (C/G), \textit{Element Recall} (C/E), \textit{Precision of Execution} (CS/GS), and \textit{Execution Recall} (CS/ES), as defined in Table~\ref{tab:metricas}.

In most test cases, the test scripts generated by \genia~correctly identified interface components, with an average of 77\% for both element metrics. Nine out of twelve test cases scored at least 70\%, and five reached 91\% or more. The exception was WebApp1-TC5, with only 12\%, revealing difficulties with dynamic, context-dependent flows. Despite this outlier, the tool showed consistent performance in standard interface scenarios.

Execution metrics were higher overall, with averages of 82\% and 85\% for relative and absolute metrics, respectively. These results reflect execution after minor manual adjustments (e.g., XPath or formatting fixes), discussed further in Section~\ref{sec:rq3} (RQ3). Ten of twelve scripts reached 84\% or more, including multiple with perfect execution. WebApp1-TC5 again performed poorly (24\%), largely due to issues propagated from the identification stage. Still, the tool proved capable of producing reliable, executable test scripts with minimal post-editing in typical scenarios. \\

\begin{tcolorbox}[
  colback=gray!5!white,
  colframe=black!75!black,
  boxsep=1pt,
  left=2pt,
  right=2pt,
  top=2pt,
  bottom=2pt,
  fontupper=\small,
  before skip=0pt,
  after skip=0pt,
  enhanced,
  sharp corners=south, 
  colbacktitle=black,
  coltitle=white,
  boxrule=0.5pt,
  title=\textbf{RQ1 – Correctness}
]

\genia~demonstrated high correctness in both element identification and test execution. Average correctness rates were 82\% for precision of execution and 85\% for execution recall. Most test cases achieved over 84\%, with only one outlier (WebApp1-TC5) revealed limitations in handling complex, context-dependent flows. Overall, the approach proved reliable in generating valid and executable test scripts for typical web scenarios.

\end{tcolorbox}

%\begin{tcolorbox}[colback=gray!5!white,colframe=black!75!black,title=\textbf{RQ1 – Correctness}]
%\genia~demonstrated high correctness in both element identification and test execution. Median correctness rates were 82\% for element identification and 88–90\% for execution. Most test cases achieved over 70\%, while only one outlier (WebApp1-TC5) revealed limitations in handling complex, context-dependent flows. Overall, the approach proved reliable in generating valid and executable test scripts for typical web scenarios.
%\end{tcolorbox}

\subsubsection{RQ2 – Completeness} \label{sec:rq2}

Completeness reflects the tool’s ability to generate test scripts that cover all interface elements and steps originally expected in the test scenarios. To assess this dimension, we used two metrics: \textit{Element Coverage} (G/E) and \textit{Step Coverage} (GS/ES), as defined in Table~\ref{tab:metricas}.

\genia~ achieved 100\% element coverage in all test cases. This indicates that the model was able to correctly infer the scope of each case, even when some elements were not extracted with full semantic precision, as discussed in RQ1.

Step coverage exceeded 100\% in several cases, with an average of 104\%. This behavior was mainly due to two complementary factors: (i) decomposition of compound steps into finer actions, such as separating the entry of email and password into two distinct interactions, and (ii) inclusion of verification steps not explicitly stated in the scenario (e.g., confirming page load after navigation). These additions did not compromise the logic of the scripts and may even contribute to more robust test validation. The low standard deviation (8.04) suggests that this behavior was consistent across test cases. Overall, the tool preserved scenario intent while enhancing test coverage through more detailed execution. \\

\begin{tcolorbox}[
  colback=gray!5!white,
  colframe=black!75!black,
  boxsep=1pt,
  left=2pt,
  right=2pt,
  top=2pt,
  bottom=2pt,
  fontupper=\small,
  before skip=0pt,
  after skip=0pt,
  enhanced,
  sharp corners=south, 
  colbacktitle=black,
  coltitle=white,
  boxrule=0.5pt,
  title=\textbf{RQ2 – Completeness}]
\genia~achieved 100\% element coverage and 104\% average execution completeness. The increase was due to step decomposition and complementary verifications. These additions preserved script logic and were consistently observed (SD = 8.04).
\end{tcolorbox}

\subsubsection{RQ3 – Adaptation Effort} \label{sec:rq3} 

RQ3 investigates the manual effort required to make the generated scripts executable, using three metrics: \textit{Modified Lines} (ML), \textit{Lines of Code} (LOC), and \textit{Modification Rate} (MR), as defined in Table~\ref{tab:metricas}. Each line edited to fix an incorrect element or action was counted as a modification. 

The modification rate had an average of 10\%, a median of 6\%, and a standard deviation of 12.95, indicating that most scripts required minor adjustments. In general, scripts that failed to execute, due to issues like incorrect locators or missing wait commands, were corrected through small changes. WebApp2-TC2 and TC4, for example, needed only 2\% and 0\% of modifications. The most common changes included refining XPath identifiers, renaming steps (e.g., replacing ``LAUNCH BROWSER'' with ``OPEN BROWSER''), and inserting synchronization commands such as \texttt{Sleep}.

WebApp1-TC5 was the only outlier (49\%), with 83 modified lines out of 171 (49\%), reflecting the challenge of managing page-to-page context in complex flows, as discussed in Section~\ref{sec:rq4} (RQ4).

Overall, the results support the feasibility of integrating \genia~into automation pipelines with little manual effort. However, the tool could benefit from a more effective mechanism to preserve execution context, especially in scenarios involving state-dependent navigation. \\

\begin{tcolorbox}[
  colback=gray!5!white,
  colframe=black!75!black,
  boxsep=1pt,
  left=2pt,
  right=2pt,
  top=2pt,
  bottom=2pt,
  fontupper=\small,
  before skip=0pt,
  after skip=0pt,
  enhanced,
  sharp corners=south, 
  colbacktitle=black,
  coltitle=white,
  boxrule=0.5pt,
  title=\textbf{RQ3 – Adaptation Effort}]
Most test cases required less than 10\% modification, with a median modification rate of 6\%. Manual edits were typically small adjustments (e.g., XPath fixes, synchronization commands) applied to ensure script executability. One outlier (WebApp1-TC5, 49\%) revealed difficulties with context loss in multi-step flows, suggesting the need for a context-preservation mechanism.
\end{tcolorbox}

\subsubsection{RQ4 - Robustness}  \label{sec:rq4}

Robustness refers to the tool’s ability to maintain consistent performance across systems with different structures, navigation flows, and interaction models. In this study, we evaluated whether \genia~can reliably generate and execute test scripts across distinct web applications, pages, and interface elements. Using the same metrics analyzed in Sections~\ref{sec:rq1} and~\ref{sec:rq2}, we examined challenges affecting script quality, execution success, and overall consistency.

Throughout the analysis, we identified five recurring challenges that impacted the consistency of outcomes. A major issue was \underline{context-dependent navigation}, where pages required prior actions to be accessible. For instance, in WebApp1-TC5, the \texttt{register} page required prior input of information such as name and email on the \texttt{input} page; otherwise, the system automatically redirected back to \texttt{input}. Although the final script included the necessary steps to perform these interactions, element identification was conducted before the full execution flow. As a result, at the time of extraction, the \texttt{register} page could not be properly loaded, leading to incomplete mappings and the inference of non-existent elements.

Similarly, \underline{dynamically injected elements}, such as error messages displayed after user interactions, were not captured during the initial identification, creating gaps in test step coverage. This suggests the need for more advanced capture techniques aligned with user flows.

Another issue was the \underline{semantic ambiguity} of natural language instructions. Terms like ``button'' or ``message'' were interpreted without considering the underlying HTML structure, leading to XPath mismatches and failed actions. A common example occurred when the instruction ``Click the 'Login' button'' was interpreted as targeting a \texttt{<button>} element, whereas the application actually used a \texttt{<a>} tag styled as a button.

In addition, robustness was compromised by \underline{external dynamic} \underline{elements}, such as ads and pop-ups overlapping clickable areas, which prevented correct interaction. In one test case, a full-screen advertisement blocked the ``Submit'' button, causing the click on the identified element to be intercepted by the banner and redirecting the browser to another site.

Finally, the tool showed fragility in handling \underline{dynamic locators}, where auto-generated IDs or unstable attributes hindered XPath reuse across executions, particularly in applications built with frameworks like React or Angular.

Overall, while \genia~delivers consistent performance in structured web scenarios, it remains sensitive to more complex contexts. These results highlight the need for future improvements to address challenges such as ensuring state preservation, handling dynamically generated elements, and building more resilient locator strategies to expand the applicability of the approach to dynamic environments. \\

\begin{tcolorbox}[
  colback=gray!5!white,
  colframe=black!75!black,
  boxsep=1pt,
  left=2pt,
  right=2pt,
  top=2pt,
  bottom=2pt,
  fontupper=\small,
  before skip=0pt,
  after skip=0pt,
  enhanced,
  sharp corners=south, 
  colbacktitle=black,
  coltitle=white,
  boxrule=0.5pt,
  title=RQ4 – Robustness]
\genia~performed reliably in well-structured scenarios but showed sensitivity to challenges such as context-dependent navigation, dynamic element injection, semantic ambiguities, external content interference, and fragile locators. Improvements are needed to address these challenges and enhance robustness in more complex environments.
\end{tcolorbox}

\vspace{-0.8em}
\subsection{Threats to Validity and Limitations}

We acknowledge several potential threats to the validity of this study, following the classification proposed by Wohlin et al.~\cite{wohlin2012experimentation}.

\textbf{External Validity.}  
The evaluation was based on two web applications and twelve test cases, limiting generalizability. Although the systems exhibit different structures and functionalities, the results may not extend to large-scale systems, dynamic SPAs, or domain-specific applications. AutomationExercise, a public platform widely used for educational purposes, was included to approximate real-world scenarios. The limited number of scenarios involving \emph{context-dependent navigation} also poses a threat, as the scenario with the most significant issues (WebApp1-TC5) depended on prior actions. Broader assessments with context-dependent flows are needed to generalize the findings. Furthermore, the evaluated systems required relatively concise prompts; in more complex applications, larger prompts may exceed token limits, affecting script completeness. Although \genia~currently uses single prompts, it can be extended to chained prompting if needed.

\textbf{Internal Validity.}  
To ensure determinism, all generations were performed with a temperature of 0. Each script was generated and executed three times under the same configuration, and the average results were reported to strengthen reliability.

\textbf{Construct Validity.}  
All data and scripts were manually evaluated under the supervision of an experienced software testing researcher. Adaptation effort was quantified using objective indicators (number of modified lines% and estimated correction time
); however, cognitive effort and debugging complexity were not captured. Future studies could complement the evaluation with qualitative methods.

\textbf{Conclusion Validity.}  
Given the small sample size ($n=12$) and skewed distributions, statistical hypothesis testing was not applied. Descriptive statistics were used as the most appropriate approach for this exploratory stage, providing insights into both typical performance patterns and edge cases.

Additionally, the decomposition of composite steps into multiple atomic actions, such as treating ``enter email and password'' as two distinct actions, affected completeness metrics without compromising test execution. The increase in completeness rates reflects granularity rather than flaws in the approach.

\vspace{-0.85em}
\section{\NoCaseChange{Related Works}}

E2E testing automation for web applications has been extensively investigated through diverse approaches, especially those aiming to validate complete user workflows through realistic interaction sequences. These efforts often intersect with broader web testing automation research, including automatic test case generation, modeling of user interface behavior via crawling techniques, and strategies to mitigate the fragility of E2E tests. A systematic mapping study conducted by Bamsam and Mishra~\cite{balsam2024web} identified three major challenges in this domain: (i) the generation of test cases and test models; (ii) test efficiency; and (iii) handling regression testing.

Regarding test case generation, several studies propose the use of models, such as navigation or formal models, as a basis for deriving abstract test cases. These models can be constructed through automated interface exploration strategies, such as navigation modeling enriched with statistical data~\cite{tonella2004statistical}, or combinatorial generation based on form submissions~\cite{wang2009combinatorial}. Alternatively, they can be extracted from the application’s source code~\cite{deng2004testing}, or built from the analysis of real usage logs~\cite{tonella2004statistical, kallepalli2001measuring, sprenkle2013configuring}. Building such models often requires domain expertise and access to detailed system artifacts. In contrast, \genia~does not rely on artifacts from the system under test and enables the generation of test scripts directly from natural language scenarios, reducing modeling effort and facilitating broader adoption.

Beyond model-based techniques, recent work has applied reinforcement learning to web-based E2E testing, aiming to discover valid action sequences through autonomous interface exploration. For example, WebExplor \cite{zheng2021automatic} incrementally builds a finite-state automaton during test execution, using temporally constrained rewards to guide interface exploration. WebQT \cite{chang2023reinforcement}, in turn, adopts a reward model inspired by the behavior of human testers to explore interactive interfaces, increasing both test coverage and efficiency. Despite their ability to autonomously explore user interfaces, these strategies face limitations in generating valid and context-aware textual inputs. In contrast to these approaches, \genia~assumes that test scenarios are already defined in natural language and focuses on translating them into executable test scripts, avoiding dynamic exploration and reducing execution complexity.

%With the recent rise of Large Language Models (LLMs), new opportunities have emerged to support software testing activities. A survey conducted by Fan et al.~\cite{wang2024software} highlights that most current applications of LLMs remain focused on low-level testing, especially unit tests, targeting tasks such as test case generation, code debugging, and failure explanation. The study emphasizes the need to explore higher levels of testing, such as integration and system testing, and to incorporate textual descriptions, such as user stories and test scenarios, into the automation process.

Another emerging line of research explores the use of Large Language Models (LLMs) and Large Vision-Language Models (LVLMs) in E2E testing. VETL \cite{wang2024leveraging}, for instance, generates context-aware textual inputs by operating directly on visual representations of the UI, rather than DOM structures. This allows it to address limitations of prior reinforcement learning techniques.  However, like those approaches, VETL does not rely on predefined test scenarios or expected outcomes, focusing instead on autonomous interface exploration.
%broaden the scope by leveraging Large Vision-Language Models (LVLMs) to generate visual and textual interactions with web applications based on annotated screenshots and multi-armed bandit strategies. The approach aims to address limitations observed in exploration techniques based on reinforcement learning by enabling the generation of semantically richer textual inputs. Nevertheless, its focus remains on autonomous interface navigation, without relying on predefined test scenarios or expected behaviors. %While promising, such approaches, like WebQT, are centered around runtime interface exploration and do not rely on pre-defined test descriptions.

Ayli et al.~\cite{ayli2024enhancing}, on the other hand, uses LLMs to enable non-technical users to create web tests through a restricted natural language interface. Their approach relies on short textual descriptions to identify interface elements based on semantic similarity, replacing traditional selectors with a smart element location mechanism.  For instance, sentences like ``the red login button'' are interpreted by GPT-4 and mapped to DOM elements using syntactic and semantic heuristics. However, the approach is limited to the automation of individual commands and does not support the orchestration of multi-step workflows or structuring complete E2E test scenarios.

In the commercial landscape, tools such as testRigor, Testim, and Functionize leverage generative AI to support E2E test automation. Testim and Functionize mainly adopt record-and-playback strategies, where user interactions are captured and later replayed to automate regression tests. While effective for automating existing workflows, these tools offer limited flexibility when the goal is to generate tests from high-level specifications or adapt them to evolving requirements. In contrast, testRigor allows testers to write test cases in plain English. Nevertheless, it remains a proprietary tool with limited integration support for open-source frameworks such as Robot Framework or Selenium. Moreover, it only supports English-language inputs, which may limit its applicability in multilingual or non-English-speaking environments.

%Model-based approaches, reinforcement learning techniques (such as WebExplor and WebQT), and methods based on LVLMs like VETL typically rely on dynamic interface exploration during test execution. While effective in navigating and interacting with complex interfaces, these strategies often require runtime instrumentation, which can limit their applicability in early development phases and complicate integration with specification-driven workflows. In contrast, \genia~ does not depend on runtime execution or interface instrumentation, which simplifies adoption and facilitates its use from the moment test scenarios are defined.

\genia~addresses these gaps by offering a free, open-source, and accessible solution that goes beyond the execution of isolated UI commands. It is capable of interpreting and automating \textbf{complete scenarios} written in natural language, including Behavior-Driven Development (BDD) syntax, by extracting sequences of actions and validations that reflect the system's expected behavior. In addition, the \genia~ approach is language-independent and automation-framework-agnostic, allowing it to be instantiated with different natural languages and integrated into various testing technologies. In the current implementation, we used ChatGPT-4o to interpret the scenarios and generate executable scripts for the Robot Framework.
Its modular architecture makes the approach particularly suitable for teams looking to automate E2E tests without the complexity of dealing directly with the technical details of the user interface.

%Compared to Ayli et al.’s approach~\cite{ayli2024enhancing}, although both leverage LLMs and aim to lower the barrier to test automation, their focus differs significantly. Ayli et al. operate at the level of individual element identification, using short, fixed-structure sentences primarily aimed at generating commands without relying on static selectors. They use the LLM to infer HTML attributes from brief descriptions of interface elements. In contrast, \genia~ interprets complete interaction scenarios—written in natural language or BDD format—extracting sequences of actions and validations that reflect the system's intended behavior. These are then translated into structured, human-readable scripts compatible with Robot Framework. This makes \genia~ particularly suitable for teams that already work with scenario-based specifications and want to automate E2E tests without dealing with low-level interface details.

\vspace{-0.8em}
\section{\NoCaseChange{Discussion}}

The results of this study provide promising evidence that the \genia~approach can effectively automate the generation of E2E test scripts from natural language descriptions. In structured web environments, the approach achieved high levels of correctness, execution success, and coverage, with minimal need for manual adjustments, reinforcing its practical viability for integration into AI-based test automation pipelines.

However, the robustness analysis revealed that \genia~is sensitive to common challenges found in real-world systems, such as the need for context preservation, the presence of dynamically injected elements, and the use of fragile locators. These factors negatively impacted the quality and completeness of the scripts in more complex scenarios. Nevertheless, the modular architecture based on multi-level prompting offers flexibility for implementing adaptations. A possible evolution would be to incorporate semantic element identification techniques, such as those proposed by Ayli et al.\cite{ayli2024enhancing}, at Level 2 of the approach, enhancing \genia’s robustness in dynamic environments with structural variations. Additionally, exploring alternative prompting strategies at this level, such as using one-shot or few-shot prompts, may reduce the need for subsequent refinement.

Despite these limitations, one of \genia’s main benefits is its ability to significantly accelerate the E2E test creation process. By minimizing repetitive and low-complexity activities, the approach allows testers to focus on more strategic tasks, such as analyzing complex flows and designing new test cases. Moreover, by reducing the need for advanced technical knowledge, \genia~broadens access to test automation, supporting its adoption by teams facing time, budget, or expertise constraints. Thus, the approach not only optimizes the practices of experienced teams but also democratizes access to quality practices across different organizational profiles.

From a practical perspective, \genia~is particularly well-suited for applications with stable page structures and predictable navigation flows, such as portals, administrative platforms, and e-commerce systems. In more complex environments, it remains viable as a support tool to accelerate the initial creation of test scripts, with human testers intervening in more context-dependent steps. Maximizing its utility in such scenarios requires careful scenario design and guaranteed access to the application during the element extraction phase.

Although the experimental evaluation focused on free-form natural language scenarios, preliminary tests using Behavior-Driven Development (BDD) scenarios written in Gherkin syntax also achieved results comparable to those obtained with natural language descriptions. \genia~was able to correctly interpret the structured Given-When-Then format and generate coherent and executable E2E test scripts. These initial findings suggest that the approach is naturally adaptable to teams adopting BDD practices.

Future work should focus on enhancing context modeling capabilities, improving locator generation heuristics, and optimizing the approach to handle very large pages. Investigating alternative prompting strategies across different levels of the pipeline is also a promising direction.  Moreover, expanding the evaluation to a broader and more diverse set of applications and combining automated metrics with user-centered assessments could provide a more comprehensive understanding of the approach's strengths and limitations. In particular, conducting qualitative evaluations with users of varying levels of expertise in test automation can offer insights into the practical applicability of \genia. Another relevant direction is to compare \genia~with commercial tools (e.g., testRigor).  We also plan to extend \genia~into a more complete test automation pipeline that includes not only script generation but also automated test data generation. This includes the integration of complementary techniques that synthesize input values from system specifications or behavioral constraints, such as the approach proposed by Mendoza et al.\cite{mendoza2024comparative}. 

Another interesting research avenue would be to explore the impact of LLM-based automation on the testing profession, examining both productivity benefits and the potential effects on the job market and skill requirements for testers in increasingly AI-assisted scenarios.

\subsection{Practical Guidelines for Using \genia}

\genia~is particularly recommended for QA teams, developers, and organizations seeking to accelerate E2E test creation without dealing directly with low-level interface details. Teams with limited experience in test automation may also benefit from the approach, reducing barriers to adopting quality practices.

Although this study focused on E2E test automation, \genia’s ability to interpret natural language descriptions and translate them into structured interaction sequences suggests broader applicability. The approach could, for example, support web scraping tasks guided by natural language navigation instructions.

The approach is most effective in applications with stable structures and conventional navigation patterns but can be adapted to more complex scenarios with specific adjustments. To maximize results, it is recommended to carefully structure the test scenarios and ensure page accessibility during the initial element extraction phase. \genia~proves particularly effective as an accelerator of the test development process, enabling human efforts to focus on refining dynamic flows and designing new validation scenarios.
\vspace{-0.85em}
\section{\NoCaseChange{Conclusion}}

This study presented \genia, an approach to automate the generation of E2E test scripts from natural language descriptions. The evaluation demonstrated that \genia~is effective in producing correct, executable, and high-coverage scripts in structured web environments, with minimal need for manual adjustments.

Despite the positive results, we acknowledge that the study presents some limitations, particularly regarding the scope of the evaluated scenarios and the diversity of systems analyzed. Nevertheless, the findings provide a solid foundation for future investigations. Future work will focus on addressing the robustness challenges identified and expanding the evaluation to other types of systems, as discussed throughout this study.

\section*{Artifact Availability}\label{sec:artifact}

The authors declare that the research artifacts supporting the findings of this study
are available at \url{https://doi.org/10.6084/m9.figshare.28873568.v5}. Additional resources can also be accessed on GitHub at \repogenia.

\begin{acks}

This paper has been supported by
CNPq - \textit{National Council for Scientific and Technological Development} (grant 420025/2023-5), FAPESP -   Fundação de Amparo à Pesquisa do Estado de São Paulo (grant 22/03090-0).

We used ChatGPT-4o to assist with the organization and revision of the manuscript and speed up the writing of LaTeX code. We carefully examined and often corrected AI suggestions, whereby we took full responsibility for the form and content of the paper.

%We acknowledge the use os AI-base tools Grammarly and ChatGPT for the improvement os spelling, grammar, vocabulary, and style. We also utilized ChatGPT to speed up the writing of LaTeX code and to support the organization of specific parts of the manuscript. We carefully examined and often corrected AI suggestions, whereby we took full responsibility for the form and content of the paper
%Não sei se essa parte temos que omitir!
% Abaixo agradecimentos Alan
% This study was financed by the Coordenação de Aperfeiçoamento de Pessoal de Nível Superior - Brasil (CAPES) Finance Code 001; in part by  Fundação de Amparo à Pesquisa do Estado de São Paulo (FAPESP) under grant number 22/03090-0, and in part by the National Council for Scientific and Technological Development (CNPq) under grant number 420025/2023-5.
\end{acks}

\bibliographystyle{ACM-Reference-Format}
\bibliography{ref}

\begin{comment}
{https://papers.ssrn.com/sol3/papers.cfm?abstract_id=3948930

https://www.techtarget.com/searchsoftwarequality/tip/How-autonomous-software-testing-could-change-QA

https://www.perforce.com/blog/alm/ai-testing-and-machine-learning-software-testing#more-on-test-case-management

https://citeseerx.ist.psu.edu/document?repid=rep1&type=pdf&doi=0fbe1b5515e747025d950658fbc039e98b29b801

https://papers.ssrn.com/sol3/papers.cfm?abstract_id=4004324

https://www.researchgate.net/profile/Swati-Chande/publication/228095029_A_Survey_on_Generation_of_Test_Cases_and_Test_Data_Using_Artificial_Intelligence_Techniques/links/6129c8f00360302a006130e4/A-Survey-on-Generation-of-Test-Cases-and-Test-Data-Using-Artificial-Intelligence-Techniques.pdf

https://www.proquest.com/openview/f1d30e9c0f4d20600396fc8f64dfa84d/1?pq-origsite=gscholar&cbl=5444811}    
\end{comment}

\end{document}